# Transiently delocalized states enhance hole mobility in organic molecular semiconductors

Samuele Giannini*[1#], Lucia Di Virgilio[2#], Marco Bardini[1], Julian Hausch[3], Jaco Geuchies[2], Wenhao Zheng[2], Martina Volpi[4], Jan Elsner[6], Katharina Broch[3], Yves H. Geerts[4,5], Frank Schreiber[3], Guillaume Schweicher[4], Hai I. Wang*[2], Jochen Blumberger[6], Mischa Bonn*[2], David Beljonne*[1]

[1] Laboratory for Chemistry of Novel Materials, University of Mons, Mons 7000, Belgium

[2] Max Planck Institute for Polymer Research, Ackermannweg 10, 55128, Mainz, Germany

[3] Institut für Angewandte Physik, Universität Tübingen, Auf der Morgenstelle 10, 72076 Tübingen, Germany

[4] Laboratoire de Chimie des Polymères, Faculté des Sciences, Université Libre de Bruxelles (ULB), Boulevard du Triomphe, CP 206/01, 1050 Bruxelles, Belgium

[5] International Solvay Institutes for Physics and Chemistry, Université Libre de Bruxelles (ULB), Boulevard du Triomphe, CP 231, 1050 Bruxelles, Belgium

[6] Department of Physics and Astronomy and Thomas Young Centre, University College London, London WC1E 6BT, UK

*Correspondence to: david.beljonne@umons.ac.be, bonn@mpip-mainz.mpg.de, wanghai@mpip-mainz.mpg.de, samuele.giannini@umons.ac.be

#These authors contributed equally to this work

## Abstract

**There is compelling evidence that charge carriers in organic semiconductors (OSs) self-localize in nano-scale space because of dynamic disorder. Yet, some OSs, in particular recently emerged high-mobility organic molecular crystals, feature reduced mobility at increasing temperature, a hallmark for delocalized band transport. Here we present the temperature-dependent mobility in two record-mobility OSs: DNTT (dinaphtho[2,3-b:2′,3′-f]thieno[3,2-b]-thiophene), and its alkylated derivative, C8-DNTT-C8. By combining terahertz photoconductivity measurements with fully atomistic non-adiabatic molecular dynamics simulations, we show that while both crystals display a power-law decrease of the mobility ($\mu$) with temperature ($T$, following: $\mu \propto T^{-n}$), the exponent $n$ differs substantially. Modelling provides $n$ values in good agreement with experiments and reveals that the differences in the falloff parameter between the two chemically closely related semiconductors can be traced to the delocalization of the different states thermally accessible by charge carriers, which in turn depends on the specific electronic band structure of the two systems. The emerging picture is that of holes surfing on a dynamic manifold of vibrationally-dressed extended states with a temperature-dependent mobility that provides a sensitive fingerprint for the underlying density of states.**

Inorganic semiconductors often sustain delocalized mobile charge carriers. For instance, the formation of large polarons, quasiparticles consisting of a mobile charge carrier dressed with a local lattice deformation (with dimensions exceeding a lattice unit cell) has been proposed to dictate the charge transport properties in various semiconductors, including lead halide perovskites[1–3] and transition metal carbides/nitrides[4]. In the absence of impurities, charge carriers are scattered by optical and/or acoustic phonons that become increasingly populated with increasing temperature. This, in turn, leads to decreasing charge mobility with increasing temperature ($T$). For instance, in the framework of the widely used effective mass approximation, the rate of acoustic phonon scattering increases as $T^{3/2}$ while the scattering from longitudinal optical phonons follows a more complex exponential dependence with temperature.[1,4] In contrast, the charge carriers in molecular semiconductors have long been assumed to be small polarons (with a size comparable to intermolecular separation); this is because the electronic coupling ($H_{kl}$, where $k$ and $l$ represent two interacting molecules) is often comparable to (or smaller than) the reorganization energy ($\lambda$).[5,6] Indeed, in the limit $H_{kl} \ll \lambda$, charge transport proceeds by incoherent hopping between states spatially confined to single molecules. This process is expected to be thermally activated, leading to $\mu$ increasing with $T$, as predicted from classical Marcus theory.[5,6] However, (weak) power-law decay of $\mu$ with $T$ has also been reported when accounting for quantum nuclear tunnelling or in the semiclassical limit of hopping theory.[7,8] Most importantly, there is now abundant experimental evidence (*e.g.*, from charge modulation [9,10] and electron spin resonance spectroscopy [11–13]) that points to charge carrier wavefunctions being spread over several molecular units in high-mobility OSs, which speaks against a small polaron hopping mechanism.[6,14,15] This discrepancy between experimental and theoretical studies has raised questions about the actual charge transport mechanism in OSs.

Narrowing down the discussion to high-mobility low-disorder single-crystal organic semiconductors, a number of experimental investigations (by time-of-flight,[16] Hall effect,[17,18] space-charge-limited current[19], and transient photoconductivity measurements[20]) have reported increasing mobility with decreasing temperature, following a power law relation $\mu \propto T^{-n}$ with $n$ ranging from ~0.5 to ~3. The reason for the large span in $n$ values is, however, unclear. On the theoretical front, exciting progress has been made lately with the development of a unified framework, the so-called "Transient Localization Scenario",[14,21,22] that helps to understand the intriguing dichotomy between extended and localized states concomitantly contributing to the thermally accessible band spectrum of a given OS.[14,23] Localized states form preferentially at the band edges. Still, the charge carriers can undergo transient quantum delocalization owing to thermal disorder, provided that they can thermally access higher-energy extended states during these dynamic excursions. This is reminiscent of the mobility edge theory,[24] but in a dynamic energy landscape due to nuclear motion (mostly low-frequency crystal phonons involving the rigid-body motion of the interacting molecules). Fratini *et al.* formalized this picture in the so-called transient localization theory (TLT).[14,22,25,26] By assuming the relaxation time approximation,[14] the authors derived a simple analytical formula in which the mobility is directly proportional to the (square of the) carrier localization length and inversely proportional to the temperature and to the fluctuation time (which depends on the period of intermolecular oscillations). TLT has been successfully applied to reconcile many experimental features of OSs and to derive design rules for the discovery of high-mobility OSs.[26,27] When employed to compute the temperature dependence of the mobility in several high-mobility OSs (*e.g.* rubrene, BTBT, C8-BTBT-C8, TIPS-pentacene etc.),[28,29] TLT provides power-law behaviour with an exponent close to 1 ($0.8 < n < 1.2$) thus suggesting a weak dependence of the ratio between localization length and fluctuation time on temperature, though this has not been confirmed experimentally. Although TLT is a very useful theory, it is still based on a number of assumptions and provides minimal physical mechanistic insights, necessarily missing how seemingly minute changes in the chemical structure can translate into very different charge transport characteristics.

Here, by combining non-adiabatic molecular dynamics simulations, in the framework of the atomistic fragment orbital-based surface hopping (FOB-SH),[21,30] with state-of-the-art ultrafast THz spectroscopy, we show how two representative high-mobility OSs that only differ by the presence of alkyl side chains, namely DNTT and its alkylated derivative, C8-DNTT-C8, feature marked differences in their $T$-dependent charge carrier mobility. Reaching such an atomistic level of resolution in the understanding of structure-property relationships has been made possible here thanks to the use of efficient numerical approaches to solve the electronic time-dependent Schrödinger equation coupled to the nuclear motion.[21,31–33] This allows us to directly follow the evolution of the charge carrier wavefunction in time, providing details about carrier transport properties missing in other analytic theories. When it comes to the study of application-relevant nano-scale systems, mixed quantum-classical non-adiabatic dynamics, using either a fully atomistic[21,34] or a coarse-grained description of the nuclear degrees of freedom,[33,35] have proven to be extremely powerful to accurately propagate charge carriers[34–36] or excitons[37–40] in OSs. The advantage of non-perturbative algorithms is that no limiting assumptions on the actual charge carrier dynamics need to be introduced, as the coupled charge-nuclear motion is solved explicitly in real-time. The wavefunction undergoes (transient) quantum (de)localization along its time-dependent dynamics, passing from localized to more delocalized states under the influence of thermal motion.[21,36,38] Thus, methods similar to those used here can provide first-principle insights into the charge transport mechanism and its temperature dependence.[41,42] It is equally important to combine these theoretical tools with experimental techniques that provide direct access to the local conductivity of materials in a contact-free fashion, as is the case for the THz spectroscopy employed here (see Methods).

DNTT and its alkylated derivative, C8-DNTT-C8, are among the best-performing molecular OSs with experimental field-effect transistor charge carrier mobility reaching up to 8-11 cm$^2$ V$^{-1}$ s$^{-1}$ in electrical transport studies by field-effect transistors.[43–45] A number of experimental and theoretical studies have been especially devoted to DNTT,[46–49] though recently also C8-DNTT-C8 has gained attention due to its sizable and isotropic electronic couplings pattern.[45] Both materials benefit from a favourable 2D charge transport character within the herringbone layer (see **Fig. 1**a,b) and a reduced dynamic disorder.[45] All these features favour large wavefunction delocalization and high charge carrier mobilities, as confirmed in this work. Future device applications for these materials are promising as an emerging technology due to their easy and low cost processability with vapour and solution deposition. Their low thermal conductivity[50] makes them suitable for thermoelectronics. Surprisingly, only very few studies have been devoted to their temperature-dependent mobility. In DNTT, the charge carrier mobility measured in OFETs was found to be almost temperature independent,[51,52] a result apparently at odds with the band-like behaviour expected for such high-mobility semiconductor.

We report below an unforeseen relationship between the temperature-dependent mobility and the electronic structure topology of these two OSs. In both molecular materials, our combined experimental-theoretical investigations show that the carrier mobility increases as the temperature decreases from 400 to 78 K, yet at a rate that is significantly higher in C8-DNTT-C8 compared to DNTT. Our modelling reveals that such a different evolution is driven by distinctive features in their electronic band structure, namely the different relative signs of the couplings and the degree of in-plane couplings anisotropy. In DNTT, this results in relatively localized tail states at the top of the valence band featuring similar extensions at varying temperatures. In contrast, C8-DNTT-C8 shows increasingly delocalized states at the valence band edge as the temperature decreases. As a result, charge carriers can thermally access more extended states giving rise to a larger mobility in C8-DNTT-C8, in particular at lower temperatures, hence the stronger falloff of $\mu$ with $T$ (as confirmed by our THz measurements).

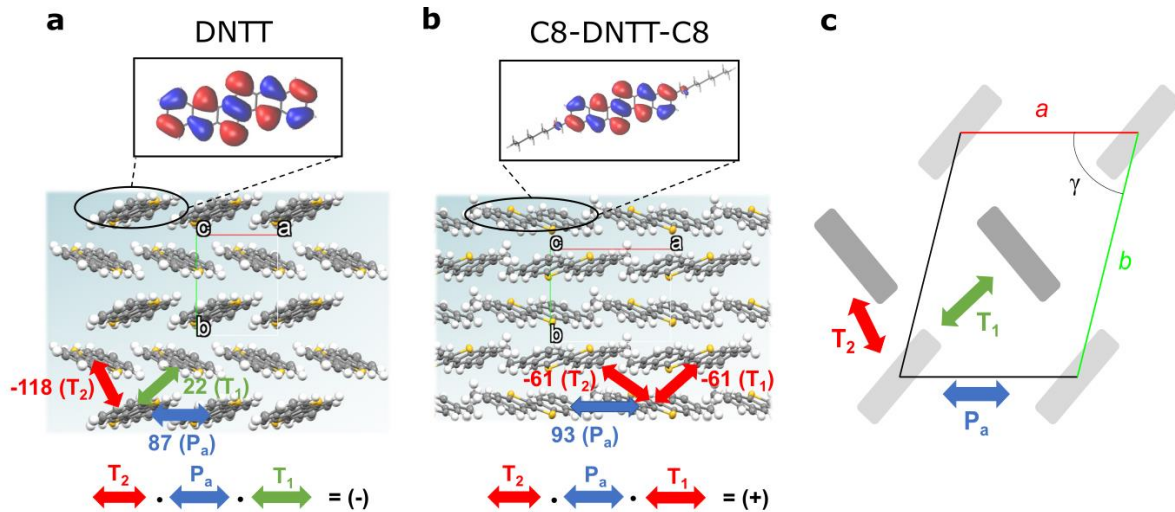

**Figure 1:** Molecular herringbone layer packing for the investigated OSs. The three largest nearest-neighbor couplings are represented in the *a-b* plane of **(a)** DNTT and **(b)** C8-DNTT-C8 (alkyl side chains have been replaced by methyl groups for clarity). The DFT highest occupied molecular orbital (HOMO) of single molecules are depicted as isosurfaces for both systems. The positive (negative) coupling-sign relationship characterizing C8-DNTT-C8 (DNTT), is represented by colored arrows and described in details in the text (same colors have been used for equivalent coupling values). **(c)** Representation of a general two-dimensional unit cell. In all panels the unit cell axes *a, b* are shown in red and green, respectively (axis *c* is eclipsed by the other two).

## Results

**Photo-induced conductivity of DNTT and C8-DNTT-C8.** To unveil the charge transport effects in OSs, we employ contact-free THz spectroscopy (with a probe length scale of 1 mm) on polycrystalline films of DNTT and C8-DNTT-C8 over 1 cm² deposited on fused silica substrate (see Supplementary Note 1 for sample morphology and structure characterization in Supplementary Figs. 1 to 3). In particular, optical pump-THz probe (OPTP) experiments allow optical injection of charge carriers by a fs laser pulse, and probing of the transient photoconductivity dynamics by a THz pulse (with bandwidth up to ~ 2 THz and a peak electrical field of $E$).[53–54] Photogenerated free charge carriers in OSs absorb THz radiation, resulting in an attenuation ($\Delta E$) of the transmitted THz field $E$. The photoconductivity $\sigma$ can be inferred from $\sigma \propto -\Delta E/E$, using the thin-film approximation.[55–56] The time evolution of $\sigma$ can be tracked as a function of pump-probe delay with sub-ps time resolution (see **Fig. 2**). The obtained $\sigma$ is proportional to the product of the density of photoexcited free charges $N$ and the mobility $\mu$, following $\sigma = Ne\mu = (N_{abs\_vol}\phi)e\mu$, where $N_{abs\_vol,}$ $\phi$ and $e$ represent the number of absorbed photons per volume, the photon-to-free-charge conversion quantum yield, and elementary charge, respectively. A detailed description of OPTP spectroscopy is included in the Methods section and Supplementary Note 2. Importantly, due to the transient nature (~ 1 ps duration) of the THz pulses, the charge carriers are driven over a short length scale (~ 10 nm), minimizing the probability of charge carriers interacting with defects, thereby making THz spectroscopy ideal for studying intrinsic local charge carrier mobility in OS films. This technique has been widely applied to understanding charge transport effects in inorganic[57] and in high mobility organic semiconductors[58,59] (in both thin-film[60] and dispersion geometries[61]).

We conducted $T$-dependent photoconductivity measurements in the range 78-300K. For a fair comparison between DNTT vs C8-DNTT-C8, we present the photoconductivity normalised to $N_{abs\_vol}$, or equivalently $\phi\mu$ (Eq. 1, see Methods), following beyond-bandgap excitation (by 3.1 eV pulses). As seen in Fig. 2b,c, for all dynamics at any given $T$, the photoconductivity in both samples builds up in ~1 ps (following optical excitations and ultrafast hot carrier relaxation), and lives for more than 1 ns (limited by the probe delay range) following a minor decay in the first 10 ps (see Supplementary Fig. 4). Two important conclusions can be drawn by comparing the photoconductivity dynamics of DNTT to that of C8-DNTT-C8. First, at room temperature (RT), C8-DNTT-C8 displays a substantially higher (by a factor of ~ 1.5) photoconductivity than DNTT (see data Fig. 2 and Supplementary Fig. 5), in agreement with the previous work by Schweicher *et al.*[45]. Second, the photoconductivity (and hence carrier mobility) of both materials increases when decreasing temperature, suggesting a band-like transport. While this $T$-dependent photoconductivity trend is barely visible in DNTT, it is very pronounced in C8-DNTT-C8 (at 78 K, the mobility is 3 times larger than at RT). Thin films of two different thicknesses (100 and 500 nm; see Supplementary Figs. 5,6) have been studied. While the signal is found to scale with the nominal thickness, a similar $T$-dependent photoconductivity dynamics is obtained in both cases (see Supplementary Fig. 7a,b).

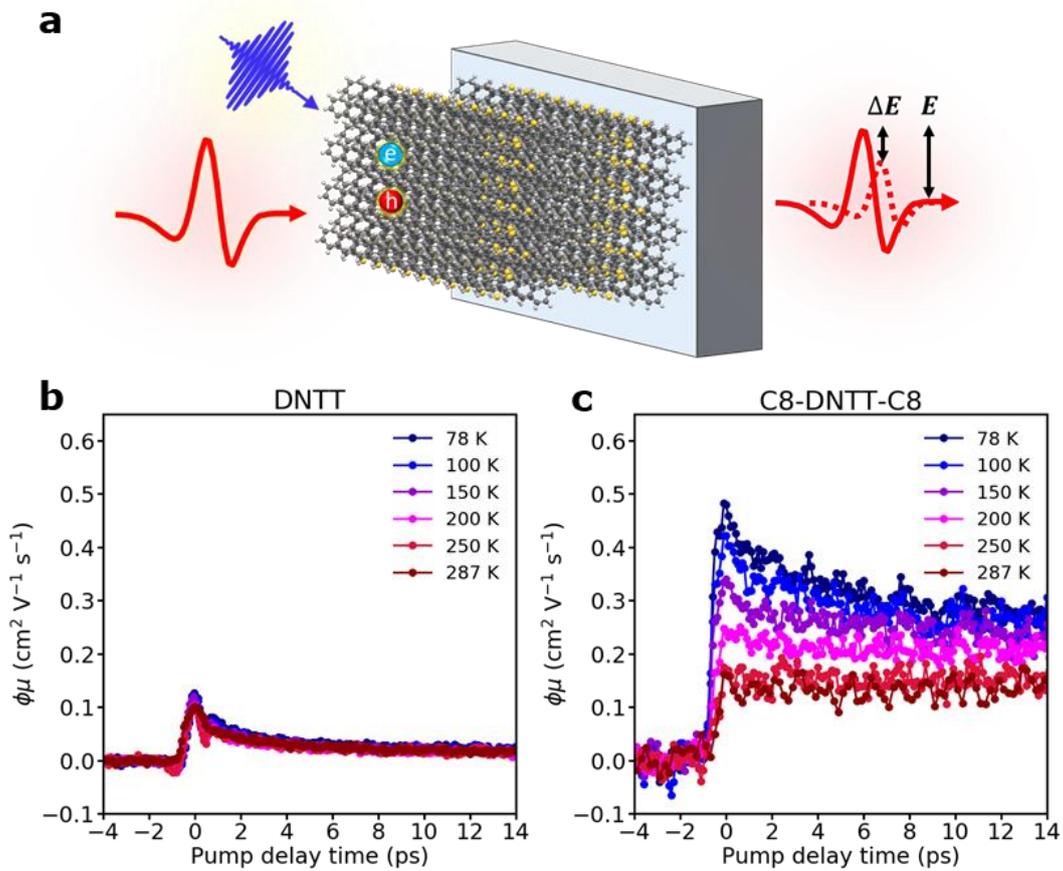

**Figure 2:** *T*-dependent photoconductivity studies by optical-pump THz-probe (OPTP) spectroscopy in DNTT and C8-DNTT-C8. **(a)** Schematic illustration of the OPTP spectroscopy. Photoconductivity dynamics of **(b)** DNTT and **(c)** C8-DNTT-C8 as a function of temperature with 3.1 eV excitation.

To obtain more detailed insights into the *T*-dependent photoconductivity trend presented in Fig. 2b,c, we recorded complex frequency-resolved photoconductivity at various *T* (at a fixed pump-probe delay of ~0.5 ps). In the *T* range studied, the photoconductivity is dominated by the real part of the conductivity (see **Fig. 3**), providing a strong indication that conduction by free carriers dominates the photoresponse in our THz bandwidth.[53] Most notably, at odds with a pure Drude-like response,[57] the real part of the photoconductivity increases with increasing frequency. Such a behaviour has been observed previously in rubrene, by both field-effect transistor and photoconductivity studies.[59,62,63] Fratini *et al.*[25,64] proposed that such a rise in conductivity with increasing frequency represents a characteristic hallmark of transient localization of the charge carrier induced by dynamic disorder.

The same authors, in Ref.[64], derived an important phenomenological model, referred to as Drude-Anderson (DA) model, which can interpolate between the Drude-like response of diffusive carriers and the finite-frequency peak expected in the presence of Anderson localization. Details of the model are given in the Method section. In brief, the DA formula, Eq. 2 (see Methods), describes transport of transiently localized charge carriers involving three different time regimes. Initially, diffusive transport of free carriers occurs following semiclassical Boltzmann theory with an elastic scattering time $\tau$. Then, a second time scale $\tau_b(>\tau)$ sets in that accounts for "backscattering" events leading to localization of charges due to molecular disorder. The dynamical nature of thermal disorder is included via a third, longer time scale, $\tau_{in}$, which is inversely proportional to the frequency of the molecular vibrations coupled to the charge carrier. The DA model describes transport of charges that are subject to localization (due to backscattering events), but that can further diffuse over a distance $L$, called localization length, with a rate $1/\tau_{in}$, owing to molecular vibrations that can trigger charge diffusion and mobility (see Eq. 4).[64] In essence, the DA model generalizes other phenomenological models used to fit the photoconductivity response by addressing localization/delocalization induced by dynamic disorder. Most importantly, DA goes beyond the Drude-Smith model[65,66] that was derived to describe charge carrier localization effects from static disorder (see a discussion of the latter model in Supplementary Note 6).

A perfectly adequate description of the photoconductivity data using the DA model, can be obtained assuming temperature-independent inelastic scattering time $h/\tau_{in}$ (see Methods and Supplementary Fig. 9a,b). From the DA fitting we find that $h/\tau_{in}$ is around 9 meV for DNTT and 16 meV for C8-DNTT-C8 in the typical range of other OSs.[22,28] These values are also in line with the related time scale found by computing the power spectral density of the coupling fluctuations (see details in Supplementary Fig. 27). The fitting procedure yields temperature-independent elastic scattering $h/\tau$ (198 meV and 300 meV) and the backscattering, $h/\tau_b$ (20 meV and 30 meV) for DNTT and C8-DNTT-C8, respectively (see Supplementary Fig. 9). Interestingly, we found that in C8-DNTT-C8 also the product $NL^2$ in Eq. 2 does not significantly change with temperature. Therefore, the increase in photoconductivity at lower temperatures is mainly a consequence of the $T^{-1}$ factor at the denominator of Eq. 3, whereas, in DNTT, $NL^2$ slightly increases with increasing temperature, and this explains the smaller increase of the photoconductivity upon lowering the temperature in this system compared to C8-DNTT-C8.

In summary, based on our THz measurements C8-DNTT-C8 is shown to possess higher photoconductivity whose amplitude further increases with lowering the temperature, much more in comparison to DNTT. The DA fitting reveals clear signs of transient localization of charge carriers in both systems. To gain a comprehensive microscopic picture, we turn to quantum-chemical calculations and full atomistic non-adiabatic molecular dynamics simulations.

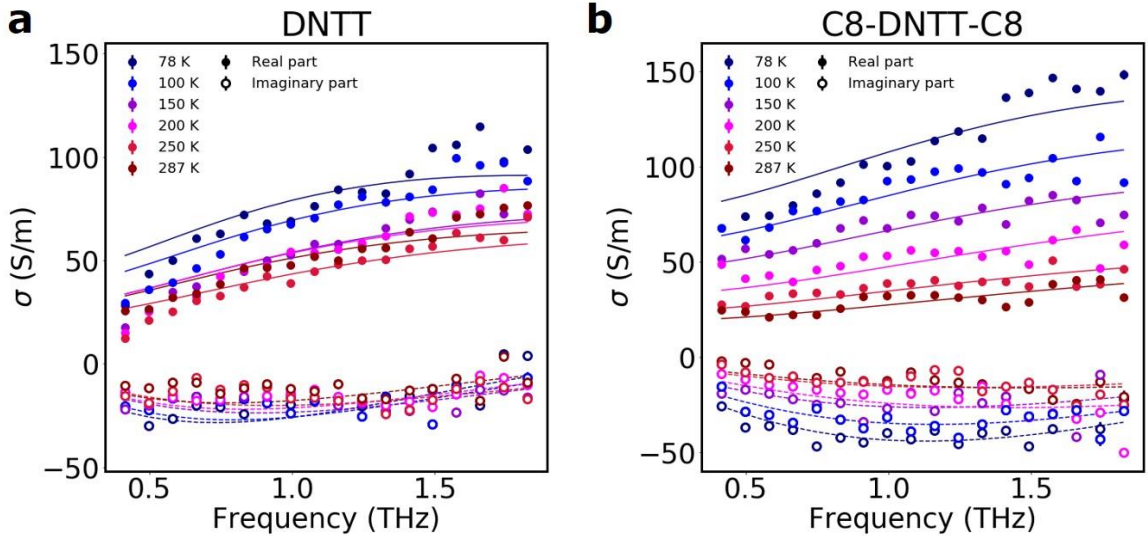

**Figure 3:** Frequency dependent THz photoconductivity. The data refer to 0.5 ps after 3.1 eV excitation of **(a)** DNTT and **(b)** C8-DNTT-C8 samples. The lines are fitting to the Drude-Anderson model described in the text (solid lines indicate the real part, and the dashed line the imaginary part of the conductivity).

**Transport parameters.** We start our computational analysis by evaluating important transport parameters, *i.e.*, internal reorganization energies ($\lambda$) and electronic couplings ($H_{kl}$) within the herringbone layer crystal structure of DNTT and C8-DNTT-C8. Cell parameters are given in Supplementary Table 1. We note at this point that the two systems show very similar unit-cell areas, which is important for a one-to-one comparison between the two (*e.g.* the absolute mobility is proportional to the square of the lattice spacing[27]). However, the presence of the alkyl side chains plays a significant role when comparing the two systems, as described in Supplementary Note 24. We refer to Supplementary Notes 8 and 9 for a description of the DFT calculations performed for reorganization energies as well as the projection operator-based diabatization (POD) and analytic overlap method (AOM) methodologies used for electronic coupling calculations. In Fig. 1a,b and Table 1, we report electronic couplings related to the transfer of hole particles. Note that in both the studied systems, holes are more mobile than electrons and should represent the majority carriers probed by our OPTP measurements (see discussion in Supplementary Note 12). We show that in both DNTT and C8-DNTT-C8, hole-transfer couplings are sizable. For some of the closest nearest-neighbour pairs, they even exceed half of the reorganization energy ($H_{kl} > \lambda/2$). In this regime, charges can delocalize over multiple molecules. Consequently, small polaron hopping model breaks down.[5,6] This is the first important observation defining high-mobility OSs.[21] Thus, alternative transport theories (*e.g.* TLT) or direct numerical approaches (*e.g.* FOB-SH) are called for. We also note that in this regime, the delocalization of the charge is limited by the strength of both local and non-local electron-phonon couplings, as clarified below.

**Table 1:** Calculated electronic couplings $H_{kl}$ for the nearest neighbor crystal pairs along different directions (Dir.), and values from literature (Lit.). All values in meV (except distances (Dist.) in Angstrom).

| System | Dir. | Dist. | $H_{kl}$(POD)[a] | $H_{kl}$(AOM)[b] | $H_{kl}$ Lit. | $\lambda$[c] |
|---|---|---|---|---|---|---|
| DNTT | $P_a$ | 6.187 | 87.4 | 80.0 | 84.8[d], 81[e] | 134 |
| | $T_1$ | 5.148 | 21.5 | 26.9 | 37.2[d], 28[e] | |
| | $T_2$ | 4.886 | -117.9 | -113.8 | -119.0[d], -94[e] | |
| C8-DNTT-C8 | $P_a$ | 5.987 | 93.1 | 83.6 | 78.9[d] | 147 |
| | $T$ | 4.941 | -60.7 | -56.2 | -60.3[d] | |

[a] Projection operator-based diabatization (POD) reference couplings are obtained as detailed in Supplementary Note 8. [b] Parametrized analytic overlap method (AOM) results are obtained as described in Supplementary Note 9. [c] Reorganization energies computed with Eq. 3 in the Supplementary Note 8. [d] All parameters refer to hole carriers. Taken from Ref.[45] (the unit-cell geometry was pre-optimized in this work[45]). [e] Taken from Ref.[47].

**Thermal disorder.** OSs, which are held together by weak van-der-Waals interactions, experience strong thermal motions of the molecules around their lattice positions. These non-local electron-phonon interactions lead to large fluctuations of the intermolecular electronic couplings (as quantified by the dynamic energy disorder $\sigma_V$). It is now clearly established that these fluctuations are detrimental to charge carrier transport in high-mobility OSs, because they contribute to scattering events which decrease mobility.[67] To explicitly account for dynamic thermal disorder, we evaluate the time-dependent Hamiltonian (Eq. 5) along FOB-SH trajectories by explicitly calculating both diagonal and off-diagonal elements on-the-fly, see Supplementary Table 6. Interestingly, we find that both DNTT and C8-DNTT-C8 show relatively small coupling fluctuations compared to their mean values ($V = \langle H_{kl} \rangle$), especially in $P_a$ direction (where $\sigma_V$ is about 4-5 times smaller than $V$). The small thermal disorder partly explains the large mobilities in these materials.[45] Coupling series as a function of time at various temperatures were also calculated as described in Supplementary Figure 27. These coupling series can be Fourier transformed to extract relevant time scales ($h/\tau_{in}$). In Supplementary Figure 27, we show that for both DNTT and C8-DNTT-C8 the power spectra peak in the energy range 3-16 meV for different coupling directions, independently of the simulated temperature (in agreement with our experimental findings). We also find that site energy fluctuations, as well as electronic coupling fluctuations, increase with increasing temperature ($\sigma_{\Delta E(V)} \propto \sqrt{T}$) as expected for an increased thermal disorder, in line with the situation in other OSs.[68] In particular, thermal fluctuations of the site energies fulfil well the expression $\sigma_{\Delta E} = \sqrt{2 k_B T \lambda}$, which is exact in the limit of linear response theory.

**Coupling-sign relationship and band structure anisotropy.** Within the transient localization framework, it was recently shown that there are two additional important parameters that govern carrier delocalization and dynamics, namely, the relative coupling-sign relationship along different crystallographic directions and the degree of anisotropy of the band structure of the material.[26] It was observed that in molecular semiconductors characterized by a 2D herringbone layer packing (which is common to the vast majority of OSs[27], including C8-DNTT-C8 and DNTT), the shape of the density of states (DOS) as well as the degree of localization of the states at the top of the valence band (or at the bottom of the conduction band) is intimately related to the sign combination and relative magnitude of the three largest nearest-neighbour electronic transfer integrals. For holes, a positive product of "signed" nearest-neighbour couplings (in the following referred to as "positive coupling-sign relation"), in combination with isotropic electronic couplings (*i.e.*, similar in magnitude), yield large carrier delocalization and, thus, fast hole carrier transport. The opposite is valid for electron transfer systems. It is worth noting that the same conclusions can be drawn from explicit numerical propagation of the wavefunction using FOB-SH non-adiabatic dynamics, as described below.[21,34]

In this respect, C8-DNTT-C8 features a positive coupling-sign relationship when considering the nearest-neighbour hole transfer couplings (see Fig. 1b), and exhibits only weak anisotropy in electronic couplings within the conductive herringbone layer. In contrast, DNTT yields an unfavourable combination (see Fig. 1a) with a negative coupling-sign relation for hole transfer (*i.e.*, a negative product of signed nearest-neighbour couplings) and a high degree of anisotropy. The consequences on the delocalization of the states of these two materials (which we will discuss in greater detail below) are already visible when computing the DOS for the frozen crystals at 0 K (black line in Supplementary Fig. 17). To this end, the one-particle electronic Hamiltonian in Eq. 5 was constructed and diagonalized, with site energies set to zero and couplings calculated at the equilibrium crystal geometry (Table 1). We note in passing that the DOS from the one-particle Hamiltonian agrees very well with the one from standard DFT band structure calculations with regards to both peak positions and bandwidth (see Supplementary Fig. 17). Interestingly, while the DOS of DNTT peaks at the top of the valence band, it instead peaks at the bottom of the corresponding band in C8-DNTT-C8. This is a straightforward outcome of the relative phase of the interacting hole wavefunctions in the herringbone plane, but one that has dramatic consequences on the energy-dependent spatial extension of the states and the associated changes in mobility with temperature, as we discuss below.

**Delocalization of the states.** The time-dependent electronic Hamiltonian (Eq. 5) computed at each time-step along non-adiabatic dynamics trajectories can be diagonalized to investigate the effect of temperature and coupling-sign relationship on both the DOS and the localization of the valence band states in DNTT and C8-DNTT-C8. Looking at the top panels in **Fig. 4**, we can see that the DOS of both materials increasingly broadens with increasing thermal disorder, *i.e.,* a larger number of tail states appear at higher temperatures (see also Supplementary Fig. 17 for a comparison of different temperatures). In the bottom panels of Fig. 4, we superimpose the DOS with the inverse participation ratio $IPR_i$, defined in Eq. 8, averaged over trajectory and time-steps for a given state, $i$, yielding $\langle IPR_i \rangle_C$. This quantity is related to the number of molecules over which state $i$ is delocalized over. The larger $\langle IPR_i \rangle_C$, the more delocalized (on average) the states that the charge carrier can thermally access. Thus, this figure depicting an IPR-resolved DOS, gives information about the spatial extent of the valence band states as a function of their energy. In DNTT (Fig. 4a-c) we can observe that, at all temperatures, the top of the valence band is formed by a dense manifold of relatively localized states that can be thermally accessed by the charge carrier wavefunction. This delocalization can be quantified by a Boltzmann average of the states $IPR_i$, $\langle IPR \rangle_B \approx 30$ at 300 K (horizontal dashed red line). The most delocalized states, which are thermally inaccessible around RT, are instead localized in the middle of the valence band at $E \approx -600$ meV ($\langle IPR_i \rangle_C \approx 200$ at 300 K). Strikingly, in C8-DNTT-C8 (Fig. 4d-f), the pattern of states (de)localization is inverted. Thermally accessible states with a very high delocalization are now found at the top of the valence band, within a few $k_B T$ of the band edge (*e.g.*, $\langle IPR \rangle_B \approx 80$ at 300 K, horizontal dashed red line). A dense manifold of localized states is instead predicted at the bottom of the valence band. This is a remarkable consequence of the sign combination, which is favourable (positive coupling-sign relation) for C8-DNTT-C8 but not for DNTT.

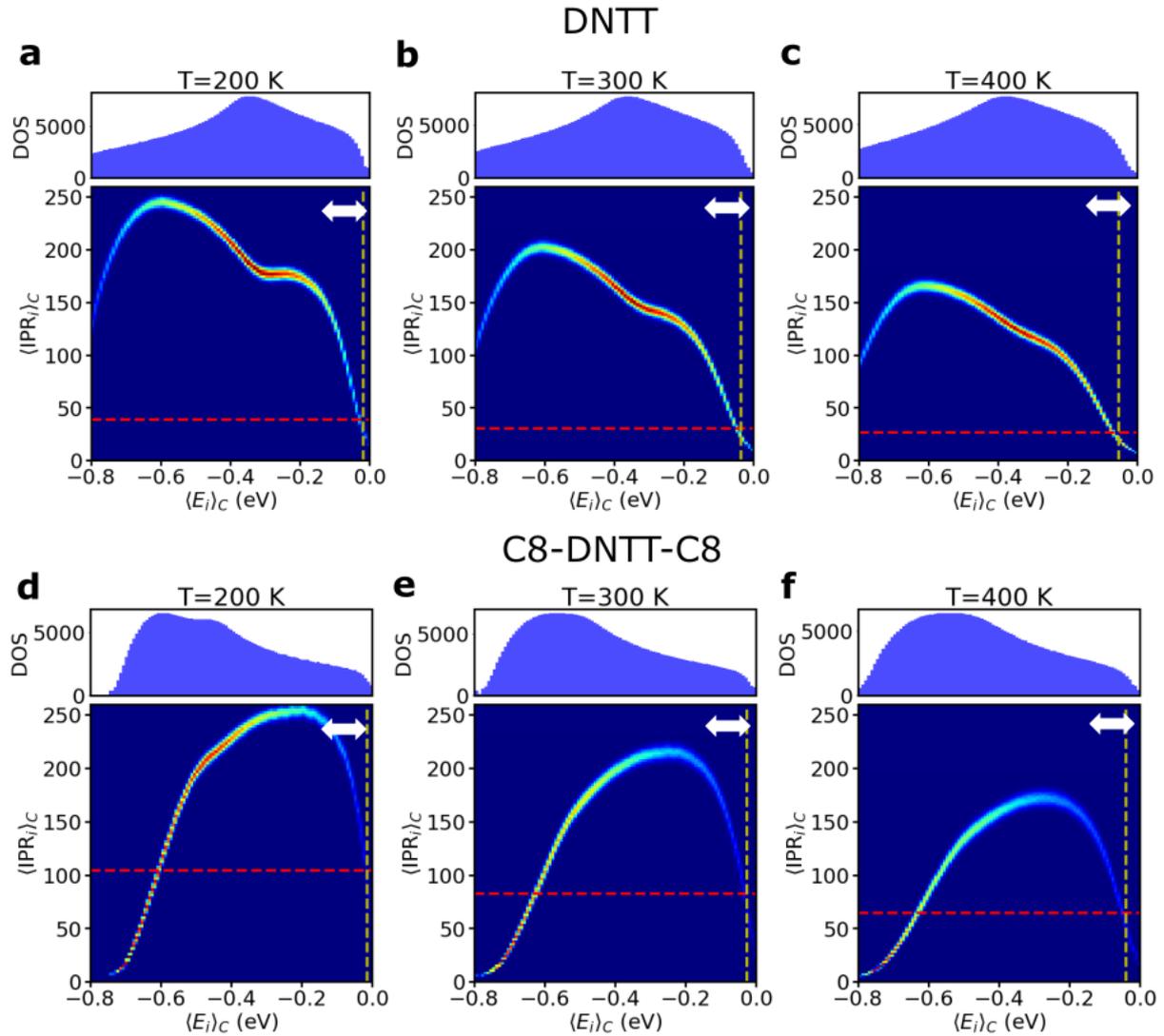

**Figure 4:** DOS and state-resolved inverse participation ratio ($\langle \text{IPR}_i \rangle_C$). Panels **(a),(b),(c)** refer to DNTT, while **(d),(e),(f)** to C8-DNTT-C8. Top figures depict the DOS at different temperatures (left to right: 200 K, 300 K, 400 K). Bottom panels display 2D histograms correlating the delocalization of the valence band states, quantified by the inverse participation ratio $\langle \text{IPR}_i \rangle_C$ (Eq. 5) versus their energies ($\langle E_i \rangle_C$). The states become denser going from regions colored in light-blue, to yellow to red (where the states are more concentrated). DOS and state-resolved IPR ($\langle \text{IPR}_i \rangle_C$) are computed from Hamiltonians extracted from around 200 FOB-SH trajectories (they include the effect of thermal disorder). Vertical dashed yellow lines indicate band active state energy $a$ ($\langle E_a \rangle_C$) which increases with increasing thermal energy. Horizontal dashed red lines are used to indicate the Boltzmann average IPR of the valence band states ($\langle \text{IPR} \rangle_B$). Note how in DNTT the Boltzmann average IPR is weakly dependent on the temperature, while C8-DNTT-C8 it decreases comparably more strongly with increasing temperature.

**Temperature-dependent charge carrier mobility.** We now turn to the main outcome of this work, namely the comparison between the simulated and measured temperature dependence of the mobility in DNTT versus C8-DNTT-C8. To this end, based on the Drude-Anderson model, we extracted the dc conductivity from frequency-resolved conductivity spectra, and the related mobility as described in the Methods section. Our analysis (see magenta lines in **Fig. 5** a,b, for DNTT and C8-DNTT-C8, respectively) reveals that the mobility follows a power law dependence, $\mu \propto T^{-n}$, with $n = 0.5 \pm 0.1$ in DNTT and $n = 1 \pm 0.1$ in C8-DNTT-C8. According to the DA model, such power laws are a consequence of the joint effect of the factor $T^{-1}$ in the mobility expression in Eq. 4 and the temperature dependence of the diffusion coefficient ($D$), related to $L^2$ by $D \cong \frac{L^2}{2\tau_{in}}$. In Supplementary Fig. 10, we show that while $D$ is roughly temperature independent in C8-DNTT-C8, yielding an overall $T^{-1}$ mobility temperature dependence, the same $D$ in DNTT is weakly thermally activated, explaining the smaller power law exponent for $\mu$ in DNTT compared to C8-DNTT-C8. From this analysis, we also found that charges in C8-DNTT-C8 travel larger distances as they are characterized by a squared localization length, $L^2$, of about 600 Å$^2$. In contrast, for DNTT, $L^2$ is about 400 Å$^2$ at RT (see Method section for details).

To confirm the previous (model dependent) DA fitting results, we directly use the photoconductivity data normalized to the absorbed photon density, $\phi\mu$ (Eq. 1), already shown in Fig. 2b,c, to confirm the trend of the mobility as a function of temperature. As depicted in Supplementary Fig. 8, our analysis unveils that $\phi\mu$ also follows a power law dependence, with $n = 0.26 \pm 0.05$ in DNTT and $n = 0.77 \pm 0.08$ in C8-DNTT-C8, respectively. Thanks to the $T$-invariance of $\phi$ (see discussion in Supplementary Table 1), the $T$-dependent scaling of $\phi\mu$ predominantly originates from the change in $\mu$. This independent measure of the temperature-dependent mobility thus provides results fully consistent with the DA method.

To uncover the underlying charge transport mechanism for DNTT and C8-DNTT-C8 and to rationalize the experimental temperature dependent data, we run a swarm of FOB-SH non-adiabatic dynamics trajectories at various temperatures. FOB-SH is used in combination with some important extensions of the original surface hopping method[69] that are necessary for accurate dynamics: decoherence correction, removal of decoherence correction induced artificial long-range charge transfers, tracking of trivial surface crossings, and adjustment of the velocities in the direction of the nonadiabatic coupling vector in the case of a successful surface hop (see details in Methods).[21,30,70] The mean squared displacement (MSD) of the wavefunction $\Psi(t)$ is evaluated with (Eq. 11), and we also quantified the delocalization of $\Psi(t)$ using the average IPR (Eq. 9), as done in previous works.[34,70] From the calculated MSD we can extract first the diffusion coefficients and then the mean plane values $\mu_{(ab)} = (\mu_a + \mu_b)/2$ of the charge carrier mobility for both DNTT and C8-DNTT-C8, using the Einstein relation (Eq. 10).[34,70] The diffusion coefficient and the mobility are reported as a function of temperature in Supplementary Fig. 10a,b and Fig. 5a,b, respectively, for both systems. FOB-SH simulations were conducted by following the computational protocol established in Ref.[34,36,70]. Full simulation details are given in the Methods section. The computed values were evaluated in the temperature range between 150 K and 400 K, where nuclear quantum effects are expected to be relatively small (specifically, since electronic couplings are large and tunneling barriers along the main transport directions are not present).[71] Focussing first on RT mobility values, we can see that the mobility in C8-DNTT-C8 is about 3 times higher than in DNTT. This observation is in good agreement with OPTP results (see Fig. 2 and Fig. 5), in which the alkylated DNTT is ~1.5-2 times more conductive than the native

DNTT, and also in line with previously reported literature RT mobility values in Ref.[45] Additionally, as shown in Supplementary Fig. 25, for DNTT, the degree of transport anisotropy in the $a$ and $b$ crystallographic directions given by our simulations ($\mu_a/\mu_b \approx 1.5 - 2.0$, depending on the temperature) is consistent with previous theoretical estimates[47] as well as other experimental OFETs mobility measurements ($\mu_a/\mu_b \approx 1.3 - 1.7$).[51]

Remarkably, non-adiabatic dynamics simulations generate $T$-dependent trends for the diffusion coefficients (Supplementary Fig. 10a-b) and mobility values (Fig. 5a-b) of both investigated systems in line with our experimental measurements. FOB-SH confirms our previous observation that while C8-DNTT-C8 features an almost temperature independent $D$, DNTT shows a weakly activated $D$. Importantly, computed mobilities (Fig. 5) in these two systems scales with power-law exponents of $n = 0.38 \pm 0.08$ and $n = 0.83 \pm 0.06$ for DNTT and C8-DNTT, respectively. The agreement between our calculations, the experimental photoconductivity results, and previous literature data is adequate. The remaining differences, *e.g.*, the discrepancies in the computed power-law with respect to temperature and the absolute mobilities compared to experiment could be due to several factors not included in the model, such as the presence of residual static defects or molecular misalignment, as well as inherent limitations of the computational method. Despite these residual uncertainties, both experiment and computation agree on the higher RT mobility (*i.e.*, factor ~2.5) and the steeper power-law scaling of C8-DNTT-C8 vs. DNTT. We discuss the physical origin of these effects in the following.

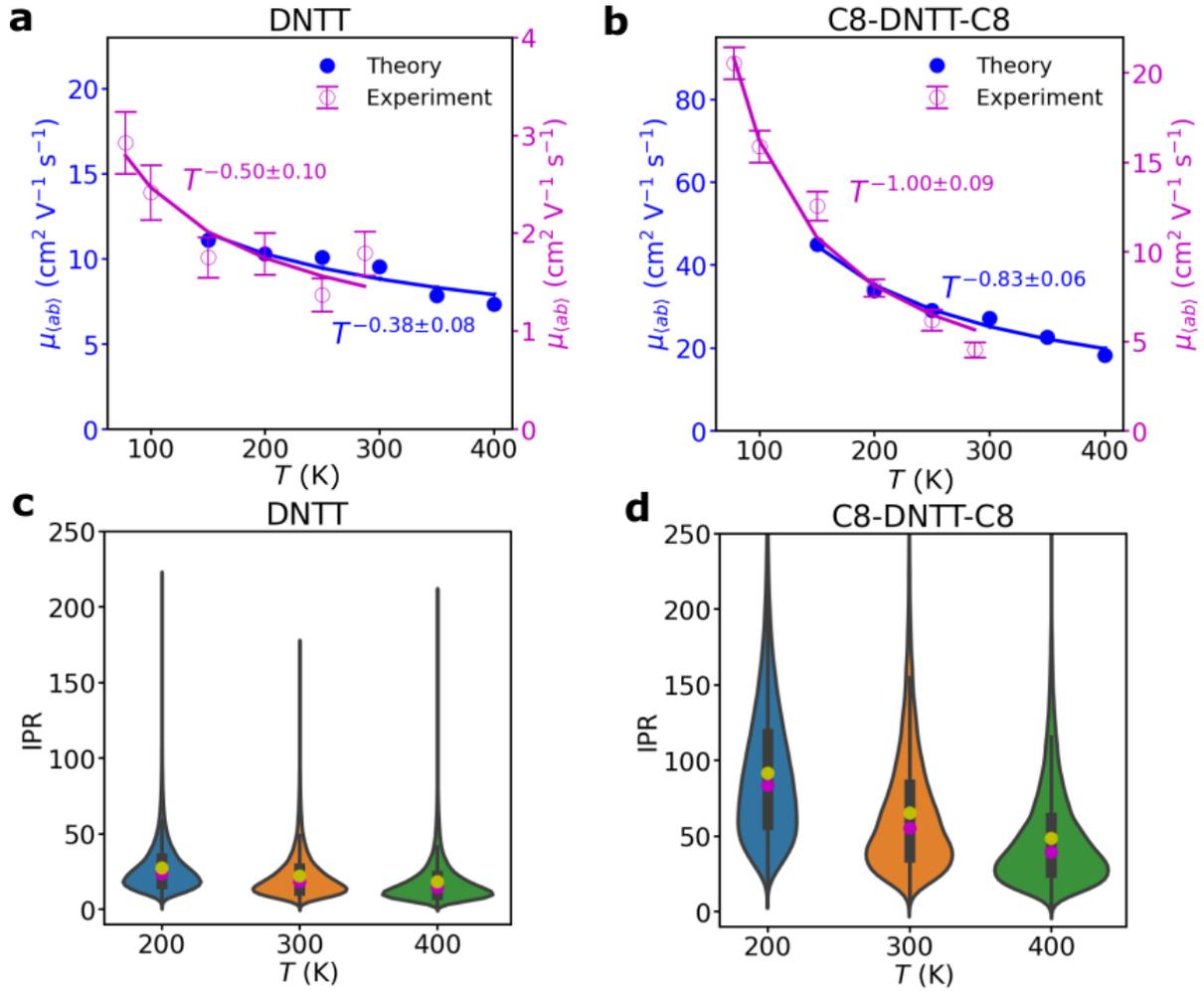

**Figure 5:** Temperature dependent IPR and experimental and theoretical charge mobilities for DNTT and C8-DNTT-C8. **(a)** and **(b)** Experimental μ, obtained from Drude-Anderson model (magenta data) and theoretical μ of hole (blue data) as a function of temperature and related power-law fitting. Error bars represent the error propagated from the standard error on the fitting parameters (details in the Method section). Theoretical mobilities are given as an average over the *a* and *b* crystallographic directions of the herringbone layers of the two OSs as described in the text. The uncertainties obtained from the fitting is given for both theory and experiment. Panels **(c)** and **(d)** violin plots representing IPR distribution (obtained from 200 FOB-SH trajectories) as a function of temperature for DNTT and C8-DNTT-C8, respectively. Black bars in the center represent interquartile ranges, while the tinner black lines stretching from the center represent the Tukey's fences. Magenta and yellow dots represent median and mean of the distribution, respectively. The mode of the distributions can be inferred by their maximum width. Note that in C8-DNTT-C8 the IPR distributions have longer tails at all temperatures compared to DNTT and the average IPR decreases more strongly in C8-DNTT-C8 compared to DNTT, indicating a more efficient quantum delocalization in the former system.

## Discussion

Having clearly highlighted essential differences between DNTT and C8-DNTT-C8 from both OPTP spectroscopy measurements and non-adiabatic molecular dynamics simulations, we now address the question of the origin of: (i) the higher RT mobility, and (ii) the steeper power-law dependence in C8-DNTT-C8 compared to DNTT. The answer to both questions can be found in the transient quantum (de)localization mechanism mediating charge transport.

By focussing on the dynamics of the charge carrier wavefunction, $\Psi(t)$, we observe that a finite-size charge carrier wavefunction is generated under the influence of thermal disorder in both systems in our simulations. By using the average IPR of the wavefunction, $\langle IPR \rangle$ (Eq. 9), to quantify the expansion of the charge carrier, we obtain that at 300 K the wavefunction is delocalized on average over 65 and 22 molecules in C8-DNTT-C8 and DNTT, respectively (see yellow dots in Fig. 5c,d). Albeit not exactly, the average charge carrier size in both systems agrees well with the extension of the thermally accessible states available at a given temperature, which we quantify using the Boltzmann average IPR, $\langle IPR \rangle_B$ (see Method section). This can be readily explained by considering that $\Psi(t)$, which is a superposition of different eigenstates $\psi_i$ (with $i$ being the index of a given state), closely tracks the extension of the tail states at the top of the valence bands of these materials (see Fig. 4). These energetically low-lying *hole* states provide predominant contributions to $\Psi(t)$, which follows Boltzmann equilibrium statistics (in the long time limit) to a very good approximation in our algorithm.[72]

Notably, though $\Psi(t)$ is a finite-size charge carrier – on average over the entire swarm of FOB-SH trajectories – each individual trajectory reveals that the charge carrier is essentially a highly dynamical "flickering" object with the tendency to delocalize over an even larger number of molecules with respect to its average (see skewed IPR distribution in Fig. 5c,d). By following the typical evolution of the charge carrier wavefunction in time at different temperatures, we find that, in several instances (shaded regions in Supplementary Fig. 18), the $\Psi(t)$ undergoes transient (short-lived) thermal intra-band excitations that bring $\Psi(t)$ from relatively localized tail states to more delocalized states closer to the middle of the valence band (see also Fig. 4). This results in a significant expansion of $\Psi(t)$ over many molecules (in some cases, about two times the average charge carrier size). Such transient expansions of the wavefunction, by which the charge carrier can "surf" highly delocalized electronic states, drive the wavefunction displacement to longer distances. Importantly, the more extended the thermally accessible eigenstates at the top of the valence band, the higher the wavefunction delocalization and possible long-range displacement that the charge can undergo (refer to the comparison between 200 K vs 400 K in Supplementary Fig. 18). This means that when averaging over the entire swarm of trajectories, we expect that these thermally accessible delocalized states will lead to larger MSDs and mobility values. However, one should keep in mind that the access to more spatially delocalized states at lower temperatures is, of course, partly compensated by the reduced thermal energy available to surf over such states (see Supplementary Fig. 22).

To further clarify the impact of the transient (de)localization effects, we note that, in Fig. 5c,d: (i) the IPR distribution (at all temperatures) is shifted to higher IPR values in C8-DNTT-C8 as compared to DNTT and; (ii) thermally accessible delocalized valence band states, as in the case of C8-DNTT-C8, favour a much broader tail at larger IPRs forming a right-skewed distribution. These characteristics underpin the more effective transient (de)localization mechanism occurring in C8-DNTT-C8 and, thus, the larger mobility found in this system compared to DNTT, thereby addressing the first initial question. We stress again that the larger delocalization of the thermally accessible valence band states in C8-DNTT-C8 compared to DNTT is intimately related to the favourable coupling-sign relationship of nearest-neighbour electronic transfer integrals along different crystallographic directions and to the higher degree of isotropy of the band structure. To nail down the crucial effect of the relative signs of the couplings in different directions, we recomputed the IPR-resolved DOS for a hypothetical, "modified", DNTT system where all the signs of the off-diagonal elements are reversed. This allows retrieving an artificially positive coupling-sign relationship as for C8-DNTT-C8. In Supplementary Fig. 19, we compare the IPR-resolved DOS for the "real" and "modified" DNTT. We show that this modification essentially inverts the IPR-resolved DOS moving delocalized tail states at the top of the valence band in the "modified" case. This effect increases twofold the $\langle IPR \rangle_B$ of "modified" DNTT compared to the "real" system (going from ($\langle IPR \rangle_B$=30 to 55).

We now proceed to discuss the reasons behind the different mobility temperature dependence characterizing C8-DNTT-C8 and DNTT, which can be understood again in the framework of transient (de)localization. In particular, by observing the vertical coloured lines Fig. 5b,c, we can see that the average IPR of $\Psi(t)$ decreases more strongly in C8-DNTT-C8 than in DNTT with increasing temperature, qualitatively tracking the steepness of the mobility falloff with temperature in the two systems. As previously noted, in C8-DNTT-C8, the IPR distributions become skewed comparably more strongly towards higher IPR values with decreasing temperature than in DNTT. This foretells a more efficient quantum delocalization as a function of temperature in C8-DNTT-C8 (see also Supplementary Fig. 18). Remarkably, these results can be traced back to the shape of the state resolved IPR reported in Fig. 4. By zooming into the tail of the thermally accessible states at different temperatures, as shown in Supplementary Fig. 20, we can see how the slope of the top valence band-edge states (within a few $k_BT$) is much steeper in C8-DNTT-C8 compared to DNTT. In C8-DNTT-C8, this slope increases more strongly (*i.e.*, thermally accessible states become consistently more delocalized) with decreasing temperature than in DNTT. This larger delocalization partially offsets the energy penalty introduced by decreasing the thermal energy available for the carrier wavefunction to access those states (Supplementary Fig. 22). This means that the transient delocalization mechanism, ruling spatial displacements and mobility, is expected to become comparably more efficient in C8-DNTT-C8 than in DNTT upon lowering the temperature, due to the larger change in the slope characterizing the eigenstates delocalization (see black lines in Supplementary Fig. 20). Thus, we argue that a favourable coupling-sign combination, as in the case of C8-DNTT-C8, not only is important for producing extended thermally accessible states which yield higher mobilities, but also favours an increasingly steeper slope of delocalized states at lower temperature. These states remain accessible despite the lower thermal energy available with decreasing temperatures. In other words, the shape of the state-resolved delocalized density of states (which in turn depends on temperature, static and dynamic disorder, the strength of electron-phonon interactions, coupling-sign relation, etc.) arguably provides a fingerprint of the mobility temperature dependence for these two high-mobility systems.

As pointed out before, hopping models are inapplicable for both DNTT and C8-DNTT-C8 because $H_{kl} > \lambda/2$ at least in one crystallographic direction, thus there is no activation barrier for hole transfer and no small polaron formation. Moreover, hopping rate theories do not consider the important effects of the coupling-sign relationship and coupling isotropy on the band structure of the system (and related DOS). This is because hopping rates depend on the magnitude of the couplings[5,6] (roughly similar in both systems) and not on their reciprocal signs in different crystallographic directions (see Supplementary Note 18). If one ignores all of these issues and calculates (hypothetical) hopping rates, one obtains very similar average mobilities for DNTT and C8-DNTT-C8 at all temperatures (see Supplementary Fig. 23) with an inaccurate power law factor. Along similar lines, we note again that $T$-dependent mobility values obtained from semiclassical band theory are incompatible in both DNTT and C8-DNTT-C8 that do not follow the Drude model typically observed in inorganic materials (see also discussion in Supplementary Note 6). On the other hand, the DA fitting, which accounts for the transient localization of charge carriers induced by dynamic disorder, provides an overall picture in agreement with direct mixed-quantum classical FOB-SH simulations.

The considerations made in this work likely apply more broadly to other high-mobility OSs (where $H_{kl} > \lambda/2$), though the trends might be somewhat blurred by other effects (amount of thermal disorder, unit-cell area, magnitude of the reorganization energy, sample purity, etc.). In Supplementary Fig. 26, we consider the two best-known OSs, pentacene and rubrene, and their experimental temperature dependence mobilities (extracted from different experimental techniques). We can see that in pentacene, where the relative coupling-sign relationship within their herringbone planes is negative (as for DNTT) the power law factor tends to be generally smaller than for rubrene for which the coupling-sign relationship is positive (as for C8-DNTT-C8). The experimental RT mobility of pentacene is also generally lower than that of rubrene, echoing the comparison of DNTT and its alkylated derivative. These observations underline the fact that also, for this notorious pair of systems, the transient (de)localization mechanism at play is affected differently by the different underlying band structure features of the given OS (in addition to other electronic and structural properties).

Arguably all these considerations might become less relevant for OSs in which the charge is fully localized by strong local-electron phonon interactions (*i.e.*, low-mobility OSs). For these systems (in which typically $H_{kl} < \lambda/2$) the low energy band tails are dominated by localized states, and more delocalized high energy states are thermally unreachable by the charge — especially at low temperatures. In such materials, the interference effects that made the relative sign of the coupling important gradually becomes less relevant as one expects for pure hopping transport, and we speculate that other effects might become dominant. For instance, Shuai and co-workers showed that nuclear quantum effects substantially increase the transport rates, especially for systems with large reorganization energy[73]. Therefore, OSs showing a large activation energy barrier ($\Delta A^{\ddagger} = \lambda/4$) are likely subject to a comparably stronger increase of the mobility at lower temperature than at room temperature due to the possibility for the charge to tunnel through the activation barrier (tunnelling effects become less relevant at RT[71]).

In conclusion, our work provides a comprehensive description of charge transport in two record-mobility OSs, *i.e.* DNTT and C8-DNTT-C8, highlighting how seemingly small changes in chemical structure profoundly impact their $T$-dependent charge-carrier mobility. We also highlight the remarkable agreement between experiment and theory as a definite demonstration that charge transport in high-mobility molecular semiconductors proceeds through a transient (de)localization mechanism at the foundation of their different power law factors. Our work represents a step forward in reconciling the discrepancy between previous experimental and theoretical studies. In particular, we have shown that in DNTT and its alkylated derivative, which feature similar structure, lattice spacing, average couplings, and even coupling fluctuations, the mobility and, importantly, the $T$-dependence of the mobility can be significantly different. This difference can be traced to different sign combinations and degree of anisotropy of the largest nearest-neighbour couplings within the herringbone layers of these systems. While the importance of these two characteristics has been appreciated before for absolute mobilities in the context of transient localization theory, this is, to our knowledge, the first time that their relevance to $T$-dependent mobility of the charge carriers has been directly and quantitatively established using experiment and simulation.

## Methods

**Materials and sample preparation** Dinaphtho[2,3-b:2',3'-f]thieno[3,2-b]-thiophene (DNTT, Sigma Aldrich, 99% purity) was used as received. C8-DNTT-C8 was synthesized according to previously described procedures.[74] The DNTT films and C8-DNTT-C8 films were prepared by organic molecular beam deposition in an ultra-high vacuum chamber with a base pressure of $10^{-8}$ mbar. For deposition each compound was resistively heated in crucibles in Knudsen-cells and the deposition rate and film thickness was monitored during preparation by a quartz crystal microbalance (QCM), calibrated using X-ray reflectivity (XRR) measurements. DNTT films of 240nm thickness and C8-DNTT-C8 films of 100nm and 500nm thickness were prepared on fused silica substrates with a growth rate of 20 Å/min.

**Optical Pump Terahertz (THz) Probe Spectroscopy.** The details of the THz setup are described in Ref.[75]. Briefly, we operate the THz spectrometer by Ti:sapphire amplified pulsed laser system (with the following output features: 800 nm central wavelength, duration of ~50 fs and a repetition rate of 1 kHz. The THz field is generated using optical rectification in a ZnTe crystal (along <110> orientation). The THz field transmitted through the sample is measured in the time domain at a chosen delay time, using electro-optical sampling in a second ZnTe crystal. The bandwidth of the THz pulse is ~ 2 THz. Optical excitations of the OS samples are conducted by 400 nm pulses, which are generated by second harmonic generation in a BBO crystal. The conductivity of pump-induced charge carriers was studied by measuring the THz absorption induced by photoinjected charges. Specifically, we monitored the peak absorption of the THz field ($\Delta E = E_{pump} - E$) by fixing the sampling beam as a function of the relative arrival between the pump and the probe, i.e, the pump-probe delay time. We then infer the mobility as a function of pump-probe delay time, *i.e.* $\phi\mu$, by applying the thin film approximation [56] (more details can be found in Supplementary Note 2) following:

$$\frac{\sigma}{N_{abs\_vol}} = \left(-\frac{\varepsilon_0 c\,(n_{sub} + n_{air})}{l} \cdot \frac{\Delta E}{E}\right)\frac{l}{N_{abs}} = \frac{Ne\mu}{N_{abs\_vol}} \propto \phi\mu \qquad (1)$$

Where $N_{abs}$ is the absorbed sheet photon density, $n_{sub} = 1.96$ is the THz refractive index of fused silica, $n_{air}$ (~1) the refractive index of the air and $l$ the thickness of the OS films. $N_{abs}$ is obtained as the product of the incident photon density and the absorbance percentage. $N_{abs\_vol}$ is the number of absorbed photons per volume. During the temperature dependence OPTP measurements, the samples were placed inside a cryostat under vacuum conditions ($< 2 \times 10^{-4}$ mbar).

**Drude-Anderson model.** This model was derived in Ref.[64] by Fratini *et al.* to account for the suppression of the conductivity, $\sigma(\nu)$, in the low frequency range due to the presence of dynamical disorder induced by thermal intermolecular vibrations. The Drude-Anderson formula reads:

$$\sigma(\nu) = \frac{Ne^2L^2}{\tau_b - \tau}\frac{\tanh\left(\frac{h\nu}{2k_BT}\right)}{h\nu}\left(\frac{1}{1 + \frac{\tau}{\tau_{in}} - i2\pi\nu\tau} - \frac{1}{1 + \frac{\tau_b}{\tau_{in}} - i2\pi\nu\tau_b}\right) \tag{2}$$

Where $\tau$ represents the elastic scattering, $\tau_b$ is the backscattering time ($\tau_b > \tau$), introduced to describe electron localization at longer times, and $\tau_{in}$ is the inelastic time related to the slow intermolecular motion responsible for the suppression of the long-time backscattering restoring charge carrier diffusion.[64] According to this model, besides the three different time scales, the product $NL^2$ between the density of charges ($N$) and the localization length ($L^2$), constitutes an additional fitting parameter.

In Fig 3, we have fitted the photoconductivity data with the DA formula in Eq. 3 following the procedure outlined in Ref.[59]. Particularly, to limit the number of free parameters, we set $\tau/\tau_b$ to 0.1. We have checked that changing this ratio between 0.1 and 0.01 does not appreciably affect the results. We have also set $\tau_{in}$ to be a globally shared parameter at all temperature because it is a material property that can be assumed independent of temperature.[59] We verified the quality of this assumption in Supplementary Fig. 27 by explicitly computing the power spectral density of the electronic coupling fluctuations directly affected by intermolecular nuclear vibrations. We also verified that even leaving it as free parameter, $\tau_{in}$ does not change significantly with temperature. To extract $N$ from the product $NL^2$ we assumed that $L^2$ can be replaced by $L_{th}^2$ calculated from TLT simulations with Eq. 11 in Supplementary Note 23. We refer to $N$ calculated from the Drude-Anderson model in this manner, as $N_{DA}$. We estimated this value to be $N_{DA} = 1.4 \cdot 10^{18}$ cm$^{-3}$ and $3.8 \cdot 10^{17}$ cm$^{-3}$ in DNTT and C8-DNTT-C8, respectively, when averaged over all temperatures (see Supplementary Table 1). As an additional validation that $N$ remains substantially temperature independent we also calculated this value using the Drude-Smith fitting of our data ($N_{DS}$) (as detailed in the Supplementary Note 6).

The DA model[64] allows to recover the necessary parameters to calculate the dc conductivity in the limit $\nu \to 0$ as:

$$\sigma_{dc}(T) \cong \frac{Ne^2}{k_BT}\frac{L^2}{2\tau_{in}} \tag{3}$$

and so, the mobility in the dilute density carrier's regime becomes:

$$\mu = \frac{\sigma_{dc}}{Ne} \cong \frac{e}{k_BT}\frac{L^2}{2\tau_{in}} \tag{4}$$

From the comparison between Eq. 3 and Eq. 10 below, it is easy to see that the diffusion coefficient can be written as $D \cong \frac{L^2}{2\tau_{in}}$.

**FOB-SH non-adiabatic molecular dynamics of hole transport.** FOB-SH is a fully atomistic mixed quantum-classical approach that allows propagating the electron-nuclear motion in real-time for large nano-scale systems. The FOB-SH methodology has been described in detail in previous works.[21,30,34,70] Below, we only give a very brief summary of the relevant equations. As common to many OSs,[21] the valence band of DNTT and C8-DNTT-C8 is well described by the following one-particle Hamiltonian (as verified in Supplementary Fig. 17):

$$H(t) = \sum_k^M \epsilon_k(t)|\phi_k\rangle\langle\phi_k| + \sum_{k\neq l}^M H_{kl}(t)|\phi_k\rangle\langle\phi_l| \qquad (5)$$

where $\phi_k = \phi_k(\boldsymbol{R}(t))$ is the (orthogonalized) HOMO of molecule $k$ for hole transport, $\boldsymbol{R}(t)$, are the time-dependent nuclear coordinates, $\epsilon_k(t) = \epsilon_k(\boldsymbol{R}(t))$ is the site energy, that is, the potential energy of the state with the hole located at site k, and $H_{kl}(t) = H_{kl}(\boldsymbol{R}(t))$ is the electronic coupling between $\phi_k$ and $\phi_l$. The Hamiltonian in Eq. 5 represents the core of the FOB-SH method. All Hamiltonian matrix elements, that is, site energies and couplings, depend on the nuclear coordinates, which, in turn, depend on time, as determined by the nuclear (molecular) dynamics. To open up applications to large super-cell sizes, necessary to accurately compute charge transport properties in high-mobility OSs, the Hamiltonian matrix elements are calculated on-the-fly using a combination of parametrized classical force-fields for site energies[34] and a very efficient analytic overlap method (AOM)[76,77] for the computation of the electronic couplings. A full description of the technical details and reference calculations needed is given in Supplementary Note 9.

Concerning the propagation of the coupled electron-nuclear motion, FOB-SH relies on a swarm of classical trajectories which, according to Tully's algorithm,[69] approximate the evolution of a quantum wavepacket. In FOB-SH the hole carrier associated with each of the classical trajectories is described by a time-dependent one-particle wavefunction, $\Psi(t)$, expanded in the same (localized) basis that is used to represent the Hamiltonian Eq. 5,

$$\Psi(t) = \sum_l^M u_l(t)\phi_l(\boldsymbol{R}(t)) \qquad (6)$$

where $u_l$ are the expansion coefficients. The time-evolution of the wavefunction is obtained by solving the time-dependent Schrödinger equation, which, using $\Psi(t)$ Eq. 6, becomes:

$$i\hbar\dot{u}_k(t) = \sum_l^M u_l(t)(H_{kl}(\boldsymbol{R}(t)) - i\hbar d_{kl}(\boldsymbol{R}(t))) \qquad (7)$$

where $d_{kl} = \langle\phi_k|\dot{\phi}_l\rangle$ are the non-adiabatic coupling elements. The nuclear degrees of freedom are propagated according to Newton's equation of motion on one of the potential energy surfaces (PES), $\psi_a$, obtained by diagonalizing the Hamiltonian Eq. 5 and denoted as $E_a$ ($a$ indicates active surface). The nuclear motion couples to the motion of the charge carrier via the dependences on $\boldsymbol{R}(t)$ in Eq. 7, resulting in diagonal and off-diagonal electron-phonon coupling. Notably, the coupling (or feedback) from the charge to the nuclear motion is accounted for by transitions of the nuclear dynamics ("hops") from the PES of the active eigenstate $a$ to the PES of another eigenstate $j$ using Tully's surface hopping probability.[69] A detailed description of the algorithm is given in Ref.[30]. For accurate dynamics, the surface hopping algorithm also needs to be supplemented with a number of important algorithms: decoherence correction, trivial crossing detection, elimination of spurious long-range charge transfer, and adjustment of the velocities in the direction of the nonadiabatic coupling vector in the case of a successful surface hop. We refer to Refs.[30,41,70,72] for a detailed description and discussion of the importance of these additions to the original fewest switches surface hopping method.[69]

**Delocalization and mean squared displacement.** A common measure used to quantify the delocalization of a given eigenstate $\psi_i$ of the Hamiltonian in Eq. 5, is the inverse participation ratio (IPR$_i$):

$$\text{IPR}_i(t) = \frac{1}{N_{\text{traj}}} \sum_{n=1}^{N_{\text{traj}}} \frac{1}{\sum_k^M \left| U_{ki}^{(n)}(t) \right|^4} \tag{8}$$

where $U_{ki}^{(n)}(t)$ are the components of the eigenvector $\psi_i$ (*i.e.* adiabatic state $i$), in trajectory $n$ at a given time $t$. Note that $\psi_i$ can be represented in terms of the localized site basis, by $\psi_i = \sum_k^M U_{ki}\phi_k$. The numerical value of the IPR represents the number of molecules (sites) $\psi_i$ is delocalized over. In Fig. 4, this quantity is averaged over time steps (*i.e.* configurations) to give $\langle \text{IPR}_i \rangle_C$ and plotted against the energy of a given state $i$, averaged over configurations, $\langle E_i \rangle_C$. IPR$_i$ can also be Boltzmann energy weighted to give a thermal average IPR, indicated in the text as $\langle \text{IPR} \rangle_B$.

A similar definition can be used to describe the delocalization of the carrier wavefunction $\Psi(t)$, obtained by directly solving Eq. 7 along time. In this case, the IPR becomes:

$$\text{IPR}(t) = \frac{1}{N_{\text{traj}}} \sum_{n=1}^{N_{\text{traj}}} \frac{1}{\sum_k^M \left| u_k^{(n)}(t) \right|^4} \tag{9}$$

where $u_k^{(n)}(t)$ are the expansion coefficients of the wavefunction in Eq. 6 at a given time $t$, in trajectory $n$. The latter definition can be averaged over time to get an average charge carrier size ($\langle \text{IPR} \rangle$). We note at this point that Eq. 8 and 9 are equivalent only when the wavefunction of the system is a pure state (*i.e.*, an adiabatic state $i$ of the system). Although, quantum decoherence pushes the wavefunction to resemble an adiabatic state, generally it remains a superposition of several adiabatic states with different weights. Eq. 9 takes this mixing into account and is more general in characterizing the charge carrier size (and less affected by sudden changes in wavefunction character and trivial crossings[70]).

The solution of Eq. 7 gives also the possibility to compute the charge carrier mobility tensor $\mu_{\alpha\beta}$ (where $\alpha$ ($\beta$) represent Cartesian coordinates, *x, y, z*) as a function of temperature. In particular,

$$\mu_{\alpha\beta} = \frac{eD_{\alpha\beta}}{k_{\text{B}}T} \tag{10}$$

$e$ is the elementary charge, $k_B$ the Boltzmann constant and $T$ the temperature. In this work, the crystallographic directions of the plane $a$ and $b$ of DNTT and C8-DNTT-C8 (which are both monoclinic) were chosen parallel to the Cartesian coordinates $x$ and $y$. In this representation, the off-diagonal components of the mobility tensor are zero due to symmetry, and one can consider just the diagonal tensor components (along $a$ and $b$ crystallographic directions).

The diffusion tensor components, $D_{\alpha\beta} = \frac{1}{2} \lim_{t \to \infty} \frac{d\text{MSD}_{\alpha\beta}(t)}{dt}$, can be obtained as the time derivative of the mean squared displacement along the nine Cartesian components ($\text{MSD}_{\alpha\beta}$),

$$\text{MSD}_{\alpha\beta} = \frac{1}{N_{traj}} \sum_{n=1}^{N_{traj}} \langle \Psi^{(n)}(t) | (\alpha - \alpha_{0,n})(\beta - \beta_{0,n}) | \Psi^{(n)}(t) \rangle \qquad (11)$$

where $\alpha_{0,n} (\beta_{0,n})$ are the initial positions of the center of charge in trajectory $n$, $\alpha_{0,n} = \langle \Psi^{(n)}(0) | \alpha | \Psi^{(n)}(0) \rangle$.

**Simulation details.** The simulation protocol employed in this work broadly follows the one devised in our previous works.[34,36] For both DNTT and C8-DNTT-C8, a series of supercells of increasing size were built from the experimental crystallographic unit cell. The dimensions of the largest supercells constructed are summarized in the Supplementary Table 2. These supercells were equilibrated in periodic boundary conditions for the neutral state at 150 K, 200 K, 250 K, 300 K, 350 K and 400 K for 500 ps in the NVT ensemble using a Nosé-Hoover thermostat. This step was followed by at least 250 ps equilibration in the NVE ensemble. In both cases, a nuclear time step of $\Delta t = 1$ fs was used. From the NVE trajectories an uncorrelated set of positions and velocities were chosen as starting configurations for FOB-SH simulations. About a thousand molecules of DNTT and C8-DNTT-C8 within their respective rectangular region of the $a - b$ high mobility plane were treated as electronically active, that is, their HOMO orbitals were used as molecular sites for construction of the electronic Hamiltonian in Eq. 5 (see Supplementary Table 2). All other molecules of the supercell were treated electronically inactive and interacted with the active region only via non-bonded interactions. FOB-SH non-adiabatic dynamics simulations were carried out with a much smaller nuclear time step compared to the standard MD equilibration step (*i.e.* $\Delta t = 0.05$ fs for DNTT and $\Delta t = 0.1$ fs for C8-DNTT). The electronic time step for integration of Eq. 7 using the Runge–Kutta algorithm to 4th order was 5 times smaller than the nuclear time step ($\delta t = \Delta t / 5$). The small nuclear time steps used in FOB-SH were necessary to efficiently tackle trivial crossings and to achieve an accurate dynamic. Additionally, all FOB-SH simulations applied a state-tracking for an automatic detection of trivial crossings and a projection algorithm for removal of decoherence correction-induced artificial long-range charge transfer.[30,34,70] For each of both systems from about 200 to 300 classical trajectories of length 0.8 to 1 ps, depending on the size of the systems, were run to extract the MSD (Eq. 11) and related mobility values for all 5 different temperatures. For each temperature, the in-plane mobility ($\mu_{(ab)}$) was calculated and averaged over at least two different supercell sizes to reduce numerical uncertainty as much as possible (see Supplementary Table 7). Convergence of the mobilities of both systems as a function of system size has been checked and reported in Supplementary Fig. 24. The initial carrier wavefunction is chosen to be localized on a single active molecule $m$, $\Psi(0) = \phi_m$ and propagated in time according to the surface hopping algorithm in the NVE ensemble. Notably, even though the short-time relaxation dynamics of the wavefunction (which is a linear combination of the valence band states) depends on the choice of the initial condition, the long-time diffusive dynamics, diffusion constant and average IPR (Eq. 9) are the same for different initial states. This is because FOB-SH fulfils detailed balance to a very good approximation thanks to the adjustment of the velocities in the direction of the non-adiabatic coupling vector.[30,72] This essential condition, together with a decoherence correction[38], ensure that the populations of the valence band states reach thermal equilibrium at long times and avoid the infinite temperature problem of the Ehrenfest dynamics. All simulations were carried out with our in-house implementation of FOB-SH in the CP2K simulation package.[78]

## Acknowledgements

S.G. and M.Bar. would like to acknowledge Dr. Claudio Quarti for useful discussions. We are grateful to H. Burg and R. Berger for conducting scanning force microscopy (SFM) measurements. This work has received funding from the European Union's Horizon 2020 research and innovation program under the Marie Sklodowska-Curie grant No 811284. J.H., K.B. and F.S. acknowledge funding by the German Research Fundation (BR4869/4-1 and SCHR 700/40-1). G.S. acknowledges financial support from the Francqui Foundation (Francqui Start-Up Grant). Y.H.G. is thankful to the FNRS for financial support through research projects Pi-Fast (No T.0072.18) and Pi-Chir (No T.0094.22). J.J.G. gratefully acknowledges support from the Alexander von Humboldt Foundation. The work in Mons has been funded by the Belgian National Fund for Scientific Research (FRS-FNRS) within the Consortium des Équipements de Calcul Intensif (CÉCI), under Grant 2.5020.11, and by the Walloon Region (ZENOBE Tier-1 supercomputer), under grant 1117545. G.S. is a FNRS Research Associate. D.B. is a FNRS research director.
.

## Author contributions

## Competing financial interests

The authors declare no competing interests.

## Additional information

Supplementary information accompanies this paper.

Correspondence and requests for materials should be addressed to D.B. (david.beljonne@umons.ac.be), S.G. (samuele.giannini@umons.ac.be), M.Bonn. (bonn@mpip-mainz.mpg.de) and H.I.W. (wanghai@mpip-mainz.mpg.de)

## Data availability

The data that supports the findings of this study are available from the corresponding authors upon reasonable request.